\documentclass[floats,aps,prl,draft,preprint,showpacs]{revtex4}

\begin{document}
\title{Long-range order in quasi-one-dimensional conductors}

\author{S.N. Artemenko$^{1,2}$}
\email[E-mail: ]{art@cplire.ru}
\author{Thomas Nattermann$^2$}
\affiliation{$^1$Institute for Radioengineering and Electronics of
Russian Academy of Sciences,
Mokhovaya str. 11-7, Moscow 125009, Russia \\
$^2$Institute f\"ur Theoretische Physik, Universit\"at zu K\"oln,
Z\"ulpicher Str. 77, 50937 K\"oln, Germany}

\date{\today}

\begin{abstract}
We study formation of the charge-density wave long-range order in
a system of repulsive 1D electrons coupled to 3D phonons. We show
that the CDW can be stabilized by interaction with phonons in
quasi-1D crystals and semiconducting nanowires. In the case of
metallic atomic chains, interaction with phonons of 3D substrate is
not enough, and violation of the translational invariance by
commensurable perturbation or disorder is needed. Possibility of
stabilization of superconductivity in 1D electrons with attraction
by means of tunnel coupling to a 3D metal is considered.
\end{abstract}

\maketitle

The basic electronic properties of three-dimensional (3D) solids are
usually well described within Landau's Fermi-liquid picture,
including phase transitions into symmetry-broken states like the
superconducting (SC) or charge-density wave (CDW) state. Even in
cases where the bare interaction is not weak, energy and momentum
conservation reduce the relevant part of the phase space making a
perturbative approach possible. This is \textit{not} the case in
1D systems where interaction is always strong. Single-electron
quasi-particles do not exist, the only low energy excitations are
charge and spin collective modes. Such a state is called the
Luttinger liquid (LL) (for a review see Ref.~\cite{Giamarchi}). A
peculiarity of 1D systems is that long-range order (LRO) cannot
exist \cite{LL}. 
In practice however 1D systems are
embedded in a 3D world  which results in
the coupling of 1D electrons to
substrates, gates, etc..
Such systems are quasi-1D rather than purely 1D systems. Then 
LRO and phase transitions into
symmetry-breaking states are not excluded.

It was shown recently \cite{Guinea} that interaction of 1D
electrons with 3D metallic gate can lead to  
quantum phase transitions and to a
power-law decay of the order parameter at
zero temperature. We will show below that
coupling to 3D environment can result in
phase transitions and true LRO even at
\textit{finite} temperatures. Different
realizations of the 1D electrons embedded
in 3D environment will be considered. The
first one is a semiconductor-based quantum
wire in which dimensionality of the
conduction electrons is reduced by
dimensional quantization. Such wires
typically have  a diameter of the order of
10 nm, hence, they contain thousands of
primitive cells in the cross-section
perpendicular to the wire \cite{Semi}. The
phonon system in such wires can be
considered to be effectively 3D. Then
electron-phonon coupling can result in the
CDW LRO, and we will determine the
conditions under which it will occur. Such
CDW state is different from the states
characterized by quasi-LRO close to a 1D
Wigner crystal \cite{Schulz,Matveev}
expected in 1D quantum wires with
long-range Coulomb interaction.

Another interesting class of quasi-1D conductors are highly
anisotropic 3D crystals with chain-like structure. These materials
exhibit commonly metallic behavior and transitions to symmetry
broken states described in terms of Fermi-liquid ideas. For
instance, inorganic quasi-1D metals like blue bronzes, NbSe$_3$
etc. undergo Peierls' transitions to a CDW state. LL in such
metals is not formed because of the instability towards
Fermi-liquid behavior around the Fermi energy in the presence of
small inter-chain hopping integral $t_\perp \ll E_F$
\cite{FirsBraBoies}. In the same time inter-chain coupling makes
them three-dimensional and provides possibility of LRO. Whereas
the main properties of quasi-1D conductors at higher temperatures
are well understood, at low temperatures they demonstrate many
intriguing properties which are not yet explained convincingly,
e.g., 
the anomalous behavior of the dielectric function interpreted as a
new glassy phase \cite{Biljak}. 
Furthermore a transition to a low temperature state characterized
by LL like conductivity was detected in focused-ion
beam processed crystals 
\cite{ZZPM}. In
order to account for such behavior the possibility of
stabilization of the LL state by defects in quasi-1D metals was
considered \cite{AR}, but electron-phonon coupling and the
formation of CDWs was not taken into account. Though formation of
LL in quasi-1D metals is problematic because of the instability
towards Fermi-liquid behavior, LL-like behavior is expected to be
seen 
if $t_\perp$ is small
compared to the other energy scales of the system related, e.g.,
to disorder or to energy gap in the spectrum. The formation of a
CDW energy gap can indeed result from the coupling of the quasi-1D
crystals electrons to 3D phonons, as will be studied below. We
also consider a strictly 1D electronic system interacting with 3D
phonons of a substrate {(e. g., chain of metallic atoms or a
single-wall nanotube)}. We find that in this case the violation of
the translational invariance is needed to stabilize the CDW LRO.
Such a violation can be induced by commensurability effects or
disorder in the substrate.

So far we considered repulsive interaction between electrons. In
the opposite case of attractive interaction we find that in
quasi-1D systems like 1D semiconducting nanowires or quasi-1D
crystals the SC LRO can be stabilized by coupling to a normal 3D
metal, while in case of strictly 1D
electronic system 
this can be performed by a tunnel contact with a superconductor.

For brevity we consider first repulsive spinless electrons
exhibiting a 1D spectrum described by the Tomonaga-Luttinger
model. Further below we will include the modifications due to
electron spin and the long-range nature of the Coulomb
interaction. The electrons are assumed to be coupled to 3D
phonons. The system is then described by an action $ S =
S_{\textrm{el}} + S_{\textrm{ph}} +S_{\textrm{el-ph}}$ consisting
of the electronic, the free-phonon part and the electron-phonon
interaction, respectively. $S_{\textrm{el}}$ can be expressed in
bosonized form in terms of the displacement field $\Phi_\rho
(\tau,x,\textbf{n})$, where $x$ and $\tau$ denote space and
imaginary time, and $\textbf{n}$  the chain index, respectively
\cite{BarisicSchulz}:
\begin{equation}
 S_{\textrm{el}} = \sum_{\textbf{n}} \int
dx d\tau  \frac{1}{2\pi v_F}\left[ (\partial_\tau\Phi_\rho)^2 +
v_\rho^2 (\partial_x \Phi_\rho)^2\right]. \label{Sel}
\end{equation}
Here $v_\rho ={v_F}/{K_\rho}$ is the velocity of plasmons, $v_F$
the Fermi velocity, and $K_\rho$  the LL parameter measuring the
strength of the interaction: $K_\rho <1$ for repulsive and $K_\rho
>1$ for attractive interaction. Sum over $\textbf{n}$ represents
summation over conducting chains of quasi-1D crystal or over
elementary cells for semiconductor quantum wire, say, in Wannier
representation. 

Free phonons are described by the elastic
displacement  field $\varphi$ with a
harmonic action
\begin{equation}
 S_{ph} = \frac{S}{2}  \int \frac{d^3q
}{(2\pi)^3} d\tau \left[ (\partial_\tau \varphi)^2 +
\omega_{ph}^2(\textbf{q}) \varphi^2\right].\label{Sph}
\end{equation}
Here $S$ is the cross section area of the primitive cell in the
perpendicular plane, $\omega_{ph}^2(\textbf{q}) = c_\|^2 q_\|^2 +
c_\perp^2 \textbf{q}_\perp^2 $. We need to keep only slowly
varying part of the $2k_F$ component of the displacement field
leading to the CDW transition, therefore, in most of the cases the
weak dependence of $\omega_{ph}(\textbf{q})$ on $q_\|$ can be
neglected. Finally, the electron-phonon part reads
\begin{equation}
S_{el-ph}= \frac{\gamma}{ \pi \alpha } \sum_{n} \int dx d\tau
\left[ \varphi e^{2i\Phi_\rho} + \varphi^*  e^{-2i\Phi_\rho}
\right],\label{Seph}
\end{equation}
where $\gamma$ is the electron-phonon coupling constant.

A similar action was used by Voit and Schulz \cite{VSch} in the 
study of electron-phonon interaction in 1D conductors.
They found modifications of the electron spectrum and the phonon
dynamics, but no true LRO was present in the purely 1D case
considered by them. To solve the problem for the case of 3D
phonons we follow Voit and Schulz and integrate out phonon degrees
of freedom. This leaves the effective action
$S_{eff}=S_{\textrm{el}}+S_{\textrm{int}}$ with
\begin{eqnarray}\nonumber
S_{\textrm{int}}= - \frac{1}{2}\left(\frac{\gamma}{\pi\alpha
}\right)^2
\sum_{n, n_1}\,\int \limits
dx d\tau d\tau_1D_{\varphi}(\tau-\tau_1,n-n_1)\\
\cos[2\Phi(\tau,x,n) -2\Phi(\tau_1,x,n_1)]\label{eq:phonon_free}
\end{eqnarray}
where $D_{\varphi}$ is the free phonon propagator, its Fourier
transform reads $D_\varphi(\omega, {\mathbf q}_\perp) = (\omega^2+
\omega_{ph}^2({\mathbf q}_\perp))^{-1}$.

Next we consider the diagrammatic series for correlation function
$D_{\rho}(\tau, x,{\mathbf n}) = \langle \Phi_\rho (\tau,
x,{\mathbf n}) \Phi_\rho (0,0,0) \rangle$. It can be found in a
standard way in terms of free correlation function $D_\rho^{(0)}$
and mass operator $\Sigma$.
\begin{equation}
 D_\rho^{-1} =
(D_\rho^{(0)})^{-1} - \Sigma, \quad 
D_\rho^{(0)} = \pi v_F/(\omega^2 + q_\|^2v_\rho^2). \label{D}
\end{equation}
We calculate $\Sigma$ in the lowest approximation in
electron-phonon coupling, and find
\begin{eqnarray}
&&\Sigma 
= \delta (x)\left(\frac{\gamma}{
\pi \alpha}\right)^2 \Bigg[D_\varphi(\tau,n) e^{-2\langle D_{\rho}
(0,0)
-D_\rho (\tau, n) \rangle}  \label{Sigma} \\
&& \left. - \delta (\tau) \delta ({\mathbf n}) \int d\tau_1
\sum_{n_1} D_\varphi(\tau_1,n_1) e^{-2\langle D_\rho (0,0) -
D_\rho (\tau_1, n_1) \rangle} \right]. \nonumber
\end{eqnarray}

It is not simple to solve equation (\ref{Sigma}) in general case,
but we can solve it in an interesting limiting case when
temperature and phonon frequencies are small compared to
electronic energies. Note that if we neglect fluctuations of the
displacement field $\varphi$ then the action
Eq.~(\ref{Sel},\ref{Seph}) reduces to that of the sine-Gordon
model. For $K_\rho<1$ the latter describes the state with the
excitations spectrum characterized by a finite dynamic mass $M$
\cite{Giamarchi}. So we expect that the mass operator must yield a
mass $M$ in the denominator of the correlation function $D_\rho$
(\ref{D}) at frequencies $\omega \sim M \gg \omega_{2k_F} \!= c_\|
(2k_F)$. On the other hand, according to the form of
Eq.~(\ref{Seph}) the mass operator must vanish in the limit
$\omega, \,{\mathbf q}_\perp \to 0$. So we seek $\Sigma (\omega,
q_\perp)$ in the form
\begin{equation}
\Sigma (\omega, q_\perp) = -\frac{M^2}{\pi v_F} \frac{(\omega^2 +
c_\perp^2 q_\perp^2)}{(\omega^2 + c_\perp^2 q_\perp^2 + \zeta
M^2)},\label{Sig1}
\end{equation}
and assume that the parameters to be found,  $\zeta$ and $M$,
satisfy conditions $\zeta \propto (\omega_{2k_F}/M)^2 \ll 1$.
Combining Eqs.~(\ref{Sig1}) and (\ref{D}) we obtain the
correlation function
\begin{equation}
D_\rho = \frac{\pi v_F (\omega^2 + c_\perp^2 q_\perp^2 + \zeta
M^2)}{(\omega^2+\omega_1^2)(\omega^2+\omega_2^2)}.
 \label{Dro}
\end{equation}
The poles of the denominator describe two eigenmodes instead of a
single mode only in Ref.~\cite{VSch}
, the number of modes corresponding to number
of variables, $\Phi_\rho$ and $\varphi$,
\begin{equation}
\omega_1^2 \approx M^2 + q_\|^2v_\rho^2, \quad \omega_2^2 \approx
c_\perp^2 q_\perp^2 + \frac{\zeta M^2 q_\|^2v_\rho^2}{M^2 +
q_\|^2v_\rho^2}.\label{modes}
\end{equation}
The mode $\omega_1$ is related mainly to electronic fluctuations,
while the soft mode $\omega_2$ mainly controls phase fluctuations
of the new Peierls-deformed lattice, i.e. of the CDW. Note that
the infrared divergence in $\langle \Phi^2_\rho \rangle = \int
\frac{d\omega}{2\pi} \frac{S d^3q}{(2\pi)^3} D_\rho (\omega,{\bf
q})$
is removed now 
and the expectation values of the $2k_F$-component of
the electronic density is finite 
\begin{equation}
\!\! O_{CDW}= \frac{\langle e^{2i\Phi_\rho} \rangle}{2\pi \alpha}
= \frac{1}{2\pi \alpha} \left( \frac{M}{\Lambda
v_\rho}\right)^{K_\rho}
.\label{OCDW}
\end{equation}

So using Eq. (\ref{Dro}) in (\ref{Sigma}) we find that the mass
operator has indeed the form (\ref{Sigma}) provided $K_\rho$ is
not too close to 1 (\textit{i.e.} for strong enough
electron-electron repulsion) and
\begin{equation}
T, \omega_{2k_F} \ll M, \quad \ln\frac{\Lambda v}{M} \gg
\sqrt{\zeta} \ln \frac{c_\| k_F \sqrt{S} }{c_\perp}.
 \label{condition}
\end{equation}
with values of parameters
\begin{equation}
M = 2 \Lambda v_F \left(\frac{\sqrt{2 \lambda}
K_\rho}{\alpha\Lambda}\right)^{1/(1-K_\rho)}, \quad \zeta =
\frac{\omega_{2k_F}^2}{M^2}, \label{M}
\end{equation}
where $\lambda = \frac{2 \gamma^2}{\pi v_F \omega_{2k_F}^2} \ll 1$
is dimensionless electron-phonon coupling constant, and $\Lambda
\approx 1/\alpha$ is a large momentum cut-off. The last relation
in Eq.~(\ref{condition}) is easily satisfied if the anisotropy of
the phonon spectrum is not too large, but cannot be satisfied in
case of 1D phonon spectrum.

When this result is applied to quasi-1D crystals there is another
mechanism of three-dimensionality, namely, inter-chain tunnelling
that could lead to instability of the LL behavior. However, in the
presence of finite mass $M$ induced by
coupling to phonons contribution of the inter-chain hopping 
can be neglected provided the inter-chain coupling is small
enough.

The same results can be obtained by  neglecting fluctuations of
the amplitude of the 2$k_F$-displacement field, $\varphi = \eta
e^{i\chi}$, and using the self-consistent harmonic approximation
(SCHA) \cite{Giamarchi}. In this way one can calculate $\langle
\chi^2 \rangle$ 
\begin{equation}
\langle \chi^2 \rangle = \left\{
    \begin{array}{ccc}
      2 \sqrt{\zeta} K_\rho
 \int\frac{S d^2 q_\perp}{(2\pi)^2}
\frac{\omega_{2k_F}}{\omega_{ph}}  {\rm {\bf
K}}\left(\frac{\omega_{2k_F}}{\omega_{ph}}\right) & \mbox{at} & T
\ll
\omega_{ph},  \\
      \pi K_\rho \sqrt{\zeta}
\int\frac{S d^2 q_\perp}{(2\pi)^2} \frac{T \omega_{2k_F}}{c_\perp
q_\perp\omega_{ph}} & \mbox{at} & T \gg \omega_{ph}.
    \end{array}
  \right.\label{Chi}
\end{equation}
where the elliptic integral ${\rm {\bf
K}}\left(\frac{\omega_{2k_F}}{\omega_{ph}}\right) \approx  \ln
\left( \frac{4\omega_{2k_F}}{c_\perp q_\perp} \right)$ for
$q_\perp \to 0$. Thus CDW phase fluctuations, $\langle \chi^2
\rangle$, are small under conditions (\ref{condition}). If we
neglect small value $\langle \chi^2 \rangle$ then our problem is
reduced to the exactly solved sine-Gordon and massive Tirring
models \cite{VSch,Giamarchi}. The value of $M$ (\ref{M}) agrees
with these models deep in the massive region (when interaction is
strong enough). For non-interacting electrons, $K \to 1$, the
exact solutions yield the result of the classical Peierls theory
$M \sim \Lambda v_F e^{-\frac{1}{\lambda}}.$

A finite value of the order parameter in Eq.~(\ref{OCDW}) follows
from dependence of the mass operator (\ref{Sig1}) on $q_\perp$. In
a strictly 1D electronic system interacting with 3D phonons of a
substrate the dependence on $q_\perp$ is absent and $\Sigma=0$ at
$\omega, q_\perp \to 0$. This is because the phase fields $\Phi$
or $\chi$ themselves do not couple to physical fields in a
translationally invariant system. Indeed, in the limit $\omega, q
\to 0$ they describe translations of the electrons and the
lattice, respectively, as a whole. The physical fields are related
to time and spatial \emph{derivatives} of the phase fields. Hence
phase fluctuations cannot diffuse into 3D space effectively
because their interaction with 3D phonons disappears in the
infrared limit. Therefore the contribution of the mode $\omega_2$
to $\langle \chi^2 \rangle$ remains divergent and coupling to 3D
phonons does not lead to LRO similar to results of
Refs.~\cite{VSch}. To remove the infrared divergence in $\langle
\chi^2 \rangle$ one needs to break the translational invariance,
e.g., by a commensurate potential in the substrate or by disorder.

Let us first consider  the violation of translational invariance
by a potential with period $\pi/k_F$ along the $x$-direction. The
additional contribution to the action,
\begin{equation}\label{chain}
\delta S = \Upsilon \sum_{n} \int dx d\tau  [O_{2k_F}^* \varphi
(n) + O_{2k_F} \varphi^*(n)],
\end{equation}
is presented in a form resembling the electron-phonon interaction
(\ref{Seph}). Here $O_{2k_F}$ describes density perturbations,
$O_{2k_F} = |O_{2k_F}| \exp (i2k_Fx +i\chi_0)$, inducing a
potential, $\Upsilon O_{2k_F}$, that acts on $2 k_F$-phonons.

Taking into account that destruction of the LRO is controlled by
phase fluctuations one can neglect fluctuations of the amplitude
of the phonon field \cite{VSch}. Then using the SCHA we integrate
out the electronic degrees of freedom and the phonon operators at
all chains except the metallic one, $\varphi (n=0) = \eta
e^{i\chi}$. This results in an effective action for the CDW phase
\begin{equation}\label{chain1}
\delta S_{eff} = 
\Upsilon \int dx d\tau  |O_{2k_F}| \eta \,\cos (\chi - \chi_0).
\end{equation}

With Eq.~(\ref{chain1}) the results for single metallic chain
become similar to (\ref{M}-\ref{Chi}) if we substitute $c^2_\perp
q_\perp^2\eta^2$ by $\Upsilon |O_{2k_F}| \eta$. Then instead of
Eq.~(\ref{Chi}) we obtain
$$
\langle \chi^2 \rangle \sim K_\rho \sqrt{\zeta} \ln \left(
\frac{\gamma \lambda^{\frac{K_\rho }{2(1-K_\rho)}}}{\Upsilon
|O_{2k_F}| \sqrt{S}} \right).
$$
In the limit of small commensurate potential, 
$\Upsilon \ll \gamma$ or $O_{2k_F} \sqrt{S} \ll 1$ 
the logarithm can be large. However, even in this case
$\langle \chi^2 \rangle$ can be small due to the small factor
$\sqrt{\zeta}$.

The divergence in $\langle \chi^2 \rangle$ can be removed also by
disorder. In this case one has to substitute $O_{2k_F}$ in
Eq.~(\ref{chain}) by the local impurity potential  in the
substrate $V(r) = \sum_i V \delta(r-r_i)$. This leads to a term
similar to Eq.~(\ref{chain1}) but with $\sum_i \delta (x -
x_i)\cos (\chi - 2k_F x_i)$ replacing $\cos (\chi - \chi_0)$.
Though this will remove the divergence in $\langle \chi^2 \rangle$
due to quantum fluctuations,  true LRO is not expected since the
latter is destroyed by the disorder. Nevertheless, for weak
disorder the phase coherence length will be quite large as it is
in CDW conductors.

For system of spinful electrons the results are qualitatively
similar to the spinless case. The spin is taken into account by
adding to Eq.(\ref{Sel}) a spin term analogous to the charge one
\cite{Giamarchi}. The electron-phonon term (\ref{Seph}) must be
also modified \cite{VSch}, namely, the integrand is multiplied by
$\cos{\sqrt{2}\Phi_\sigma}$ where $\Phi_\sigma$ is the spin phase
field, and factors 2 in the exponents are substituted by
$\sqrt{2}$. After going through the same steps as in the spinless
case we find mass operators for correlation functions $D_{\rho}$
and $D_{\sigma}$. For the charge sector $\Sigma_\rho$ has a form
similar to Eq. (\ref{Sig1}), while $\Sigma_\sigma$ is not
dispersive, so the spin excitations are described by dispersion
relation similar to $\omega_1$ (\ref{modes}) but with different
velocity. The masses for charge and spin sectors are equal in the
main approximation
\begin{equation}
 M =  \Lambda v_F
 [\lambda \, K_\rho^{K_\rho}\,
K_\sigma^{K_\sigma}/(\alpha\Lambda)^2 ]^{1/(2-K_\rho-K_\sigma)}
 . \label{Ms}
\end{equation}
Since we consider repulsive electrons we can ignore the
backscattering in the interaction. While in case of
inter-electronic attraction it leads to the spin-gapped
Luther-Emery phase (\textit{cf}. \cite{Giamarchi}), in our problem
it results only in small corrections. 

Up to now we considered the short-range interaction and treated
$K_\rho$ as a constant. In the case of Coulomb interaction it
depends on wave vector. The exact form of $K_\rho (\mathbf{q})$
depends on crystal structure, but in the long-wavelength limit one
can describe interaction by the Fourier transform of the Coulomb
potential. Then $K^2_\rho({\mathbf q})=(1+ \kappa^2/q^2)$ where
$1/\kappa$ is 
Thomas-Fermi screening length. This results
in qualitative modifications of the spectrum of the eigenmodes.
Inserting $v_{\rho}=v_F/K_\rho$ into (\ref{modes}) we find that
the mode $\omega_1$ approaches at $q_\perp = 0$ the plasma
frequency, 
and the phason mode $\omega_2$ is not a
soft mode any more
. Similar hardening occurs in the LL density mode
without the coupling to phonons \cite{BarisicSchulz} and in the
CDW phase mode in classical Peierls theory \cite{LRA}.

Similar to the CDW case, but with action written in terms of dual
field $\Theta_\rho$ \cite{Giamarchi}, one can study stabilization
of SC LRO in a 1D electronic system with attraction, ($K_\rho >
1$). Since the SC phase is conjugate to the particle number
operator and the fixed number of particles makes fluctuations of
the phase infinite  one can try to let the particle number to
fluctuate in order to diminish fluctuations of the SC phase by
coupling to a normal or SC 3D metal via a tunnel junction
providing a way to transfer fluctuations of the phase to 3D space.
We describe the coupling to the 3D metal by a tunnel Hamiltonian.
The 3D metal is presented by the BCS Hamiltonian in the Bogolyubov
self-consistent approximation. Then we integrate out the fermions
of the 3D metal and obtain an effective action describing coupling
to the SC order parameter, $\Delta$, of the 3D metal:
\begin{equation}\label{super}
 S_{tun}^{(eff)} = \sum_{n} \int dx d\tau \, t (e^{2i\Theta}
\Delta +  e^{-2i\Theta} \Delta^*).
\end{equation}
We describe $\Delta$ by a Ginzburg-Landau free energy for a 3D
superconductor with critical temperature $T_c$. So we consider
coupling to a superconductor for $T < T_c$, and to 
a normal metal for $T > T_c$. We find that the SC LRO can be
stabilized by coupling to a normal 3D metal in quasi-1D structures
where dependence of $\Theta_\rho$ on transverse coordinates is
possible. In strictly 1D systems coupling to a normal metal does
not stabilize the LRO, and only quantum phase transition as
discussed in Ref.~\cite{Guinea} are possible. This happens
because, like in the CDW case, physical fields are related to
derivatives of the SC phase, and coupling to 3D space vanishes in
the infrared limit. So, again, the symmetry breaking is needed.
The symmetry is violated if the 3D metal is in the SC state,
$\langle \Delta \rangle \neq 0$ in Eq.~(\ref{super}), and the
problem is
reduced to the sine-Gordon model. Hence, 
the LRO becomes possible.

In conclusion, we found that coupling of electrons in quasi-1D
systems to 3D objects should lead to symmetry-breaking transitions
described in terms of the LL ideas. The transition to the CDW
state was studied in case when phonon frequency is much smaller
than typical electronic energy and the repulsion between electrons
is strong enough. Both conditions are satisfied in quasi-1D
crystals where typical values of the Peierls gap are above 500K,
and the interaction parameter $K_\rho$ is quite small, $K_\rho
\sim 0.1 \div 0.3$ \cite{AR}. Our study of the SC transition
explains possibility of the LRO formation in quasi-1D systems but
it has more model character because we consider electronic
attraction as granted and do not study mechanism of attraction.

S.N.A. acknowledges financial support by SFB 608 and by Russian
Foundation for Basic Research.

\end{document}